\documentclass{aa}
\usepackage{txfonts}
\usepackage{astro}
\usepackage{natbib}
\usepackage{graphicx}
\usepackage{amssymb}

\begin{document}
%

  \def\h{$^{\rm h}$}
  \def\m{$^{\rm m}$}
  \def\s{$^{\rm s}$}
  \def \rhop{\ensuremath{\rho \rm Oph}}
  \defcitealias{pound03}{PRB03}
  \def \prb{PRB03}

\title{Velocity field and star formation in the Horsehead
  nebula}

\author{%
  P. Hily-Blant\inst{1,2}
  \and
  D. Teyssier\inst{3,4}
  \and
  S. Philipp\inst{5}
  \and
  R. G\"usten\inst{5}
}

\offprints{P.~Hily-Blant \email{hilyblan@iram.fr}}

\institute {
  LRA-LERMA, \'Ecole normale sup{\'e}rieure et Observatoire
  de Paris, 24 rue Lhomond, 75231 Paris cedex 05, France
  \and
  Institut de Radio Astronomie Millim\'etrique, 300
  Rue de la Piscine, F-38406 Saint Martin d'H\`eres, France
  \and
  Space Research Organization Netherlands, P.O.Box
  800, 9700 AV Groningen, The Netherlands
  \and
  Departamento de Astrofisica Molecular e Infrarroja,
  Instituto de Estructura de la Materia, CSIC, Serrano 121,
  28006, Madrid, Spain
  \and
  Max-Planck-Institut f\"ur Radioastronomie, Auf dem
  H\"ugel 69, D-53121 Bonn, Germany
}

\authorrunning{P. Hily-Blant \etal}
\titlerunning{Velocity field and star formation in the
  Horsehead nebula}
\date{Received / Accepted}
\abstract{Using large scale maps in \cdo\jtwo\ and in the
  continuum at 1.2mm obtained at the IRAM-30m antenna with
  the Heterodyne Receiver Array (HERA) and MAMBO2, we
  investigated the morphology and the velocity field probed
  in the inner layers of the Horsehead nebula. The data
  reveal a non--self-gravitating ($m/m_{\rm vir}\approx
  0.3$) filament of dust and gas (the ``neck'', $\varnothing
  = 0.15-0.30\,\pc$) connecting the Horsehead western ridge,
  a Photon-Dominated Region illuminated by $\sigma$Ori, to
  its parental cloud L1630. Several dense cores are embedded
  in the ridge and the neck. One of these cores appears
  particularly peaked in the 1.2\,mm continuum map and
  corresponds to a feature seen in absorption on ISO maps
  around 7\,$\mu$m. Its \cdo\ emission drops at the
  continuum peak, suggestive of molecular depletion onto
  cold grains. The channel maps of the Horsehead exhibit an
  overall north-east velocity gradient whose orientation
  swivels east-west, showing a somewhat more complex
  structure than was recently reported by \cite{pound03}
  using BIMA CO\jone\ mapping. In both the neck and the
  western ridge, the material is rotating around an axis
  extending from the PDR to L1630 (angular velocity
  $=1.5-4.0\,\kms$). Moreover, velocity gradients along the
  filament appear to change sign regularly (3\,\kmspc,
  period=0.30\,pc) at the locations of embedded integrated
  intensity peaks. The nodes of this oscillation are at the
  same velocity. Similar transverse cuts across the filament
  show a sharp variation of the angular velocity in the area
  of the main dense core.  The data also suggest that
  differential rotation is occurring in parts of the
  filament. We present a new scenario for the formation and
  evolution of the nebula and discuss dense core formation
  inside the filament.  \keywords{ISM: clouds -- kinematics
  and dynamics -- individual objects (Horsehead nebula) --
  Stars: Formation -- Radio lines: ISM} }

\maketitle

%
\section{Introduction}
%
Dense molecular cores are the basic units of isolated
low-mass star formation. It is now common that molecular
line mapping from dark clouds reveals clumps embedded in
filamentary structures
\citep[\eg][]{onishi98,onishi99,obayashi98,loren89a,chini97}. These
filaments can be either self-gravitating or not. The clumps
are mostly gravitationally bound, and in many cases, star
formation is known to have already started. Therefore,
molecular filaments are thought to play a crucial role in
low-mass star formation.

However, little is known about the general physical
properties of these filaments. For example, the density
distribution is very likely not uniform, but instead varies
according to a power-law. \cite{stepnik03} have investigated
this observationally in a filament of the Taurus molecular
cloud, and concluded that a $r^{-2}$ profile was compatible
with the observations. Steeper density profiles can however
be observed, as is shown by Hily-Blant \& Falgarone ({\it in
prep}) in a non--self-gravitating filament connected to the
low-mass dense core L1512. Theoretical models with and
without magnetic fields suggest power-laws with exponents
ranging from $\approx-2$ \citep{fiege00a} to $-4$
\citep{ostriker64, nakamura93, stodol63}.

The importance of the velocity field in the formation of
clumps in filamentary clouds has already been noted several
years ago by \cite{loren89b}, who did a systematic study of
the velocity pattern of well-known filaments in \rhop. The
longitudinal velocity was shown to exhibit gradients near
the locations of the main clumps harboured in the
filament. From the theoretical and numerical points of view,
recent studies also stress the role of the velocity
field. Dealing with self-similar rotating magnetized
cylinders, \cite{hennebelle03} shows that the velocity field
strongly depends on the relative intensity of the toroidal
component of the magnetic fields with respect to the
poloidal one and to the gravitational force: the rotation
can be mainly that of a rigid body or instead be
differential. The importance of the velocity field has been
further stressed by \cite{tilley03} who show how it could
help in distinguishing between collapsing and equilibrium
cylindrical distributions and between magnetized and
unmagnetized filaments. \cite{fiege00a} performed numerical
studies of the fragmentation of pressure truncated
isothermal filaments threaded by helical magnetic fields and
found that the toroidal velocity could change sign
periodically when magnetic field dominates over
gravity. More recently, \cite{sugimoto04} numerically
studied the decay of Alfv\'en waves in filamentary
clouds. They show that the propagating circularly polarized
waves generate longitudinal sub-waves that steepen into
shocks where the initial energy is being dissipated. While
propagating, the circularly polarized Alfv\'en waves make
the filament rotate. In some cases, the filament can
fragment into regularly spaced clumps.

Instabilities (gravitational, magnetic, or both) are often
invoked to explain the formation of dense cores in
filaments. Nonetheless, molecular observations of
filamentary structures at high spatial and spectral
resolution are still lacking, which would allow both the
density and the velocity fields to be constrained. The
scales of interest are typically of the order on 0.1\,pc and
velocity gradients of the order on 1\,\kmspc. In this paper,
we investigate the velocity field and its link to the
density distribution in the Horsehead nebula, a close-by
(400\,pc) dark protrusion emerging from its parental cloud
L1630 in the Orion B molecular complex.  This condensation
is illuminated by the O9.5V star $\sigma$Ori (distance
0.5$^\circ$ from the cloud) and presents a Photon-Dominated
Region (PDR) on its western side seen perfectly edge-on
\citep{abergel03}. Until some years ago, molecular
observations of this source were still scarce and limited to
low spatial resolution \citep[2\arcmin\ resolution,
][]{kramer96}. New insights at higher resolution
(10--20\arcsec) have recently been revealed by
\cite{abergel03} and \cite{pound03}, who observed the
Horsehead in various millimetre transitions and isotopes of
CO. In particular, \citeauthor{pound03} (hereafter \prb)
analyse the velocity field of this source, using
interferometric CO\jone\ data
(10\arcsec$\times$8\arcsec~resolution), and reported a
general NE-SW velocity gradient of order 5\,\kmspc\ across
the nebula. They also note that star formation may have
already started in the cloud and discuss possible scenarios
that could have driven the formation of such a
structure. The use of an optically thick probe may however,
have hindered them from obtaining a thorough picture of the
structure in the innermost layers of this object. In the
present study, we make use of the rarer isotopomer \cdo, as
well as the emission of dust at millimeter wavelengths.

The paper is organized as follows. After a brief description
of the observations in Sect.~\ref{obs}, we analyst in
Sect.~\ref{spatial} the spatial distribution of the
\cdo\jtwo\ emission and of support data obtained in the
continuum at various wavelengths to derive representative
physical parameters. The analysis of the velocity field is
then explained in Sect.~\ref{velo}, and its consequences on
the genesis and evolution of the object are then discussed
in Sect.~\ref{discussion}. We finally summarize our
conclusions in Sect.~\ref{conclusion}.

%
\section{Observations}
\label{obs}
%

The \cdo\jtwo\ data presented in this paper have been
observed in May 2003 using the new Heterodyne Receiver Array
\citep[HERA,][]{schuster03,schuster04} operated at the
IRAM-30m telescope. The data were obtained in the On-The-Fly
mode with an array inclined on the sky by 18.5$^\circ$ in
the equatorial frame, providing a direct sampling of
8\arcsec\ in Declination, while a derotator located in the
receiver cabin was used to keep the HERA pixel pattern
stationary in the Nasmyth focal plane. The half-power
beam-width is 12\arcsec.  The data were obtained under
good-to-average weather conditions and system temperatures
were on the order of 400\,K, providing an average noise
r.m.s. in the regridded map of 0.2\,K ($T_{\rm A}^*$ scale)
within 0.1\,\kms\ velocity channel. The final data are
scaled in $T_{\rm mb,c}$, a temperature scale which takes
into account the emission peaked-up by the 30-m error beam
at 1.2\,mm \citep[see][and references therein for
details]{abergel03}.  Figure~\ref{hh fig} shows the
corresponding integrated intensity map in the velocity
interval $[9.1:11.8]$\,\kms.

\def\wa{0.75\hsize}
\begin{figure*}
  \begin{center}
	\includegraphics[width=\wa,angle=270]{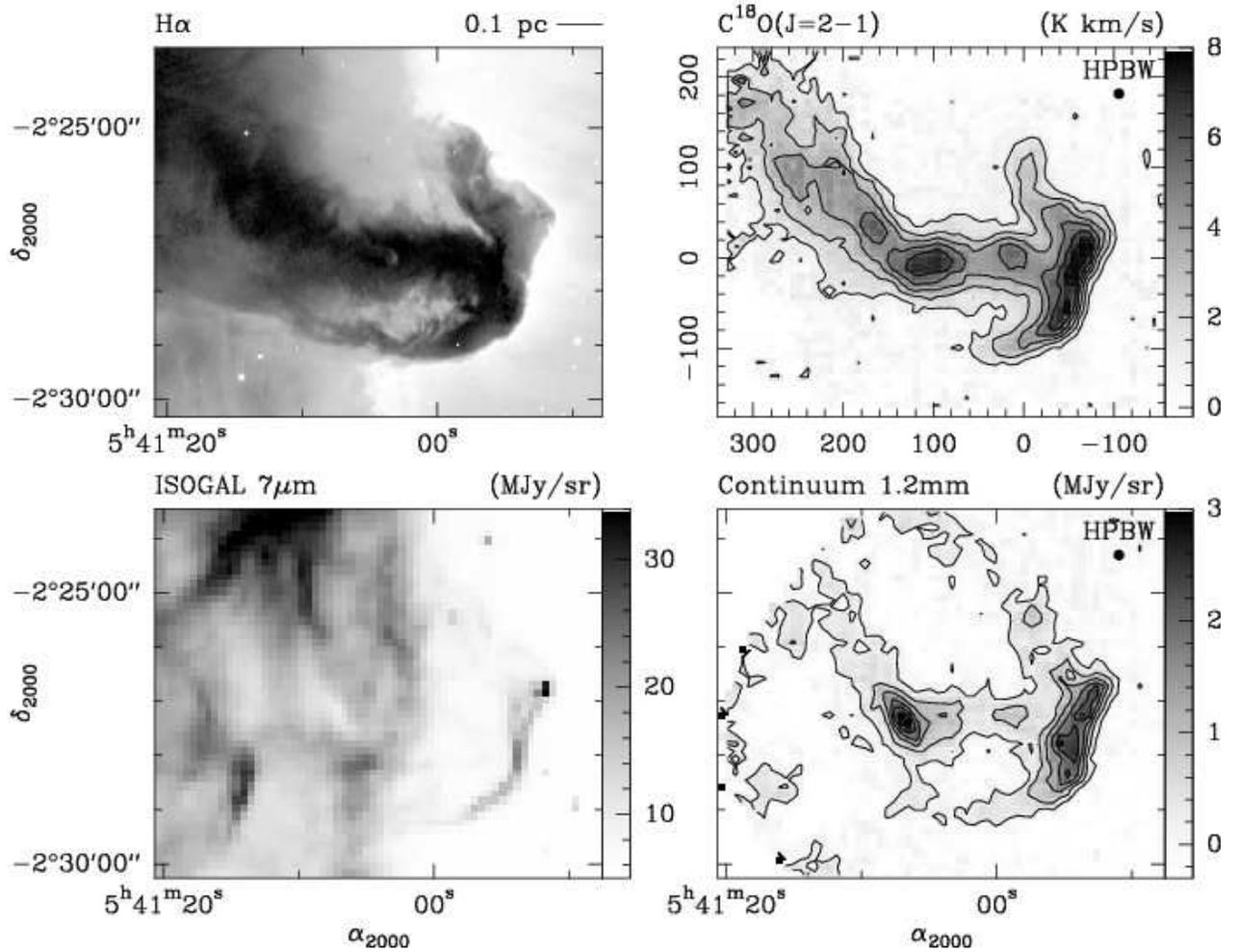}
	\caption{{\it (Upper left)}: H$\alpha$ line emission map
	  obtained at the 0.9\,m KPNO telescope (courtesy of
	  Reipurth \& Bally). {\it (Upper right)}: Integrated
	  emission map ($\int T_{\rm mb,c}\ud v$, \kkms,
	  $v\in[9.1:11.8]\,\kms$) of C$^{18}$O\jtwo.  Offsets
	  are in arcsec and contours are 1.2 to 7.2 by
	  1.2\,\kkms. The half-power beam width (HPBW) is
	  plotted at the top right corner. {\it (Bottom left)}:
	  continuum emission around 7\,$\mu$m (mostly aromatic
	  emission) covered by ISO. {\it (Bottom right)}:
	  Continuum emission at 1.2\,mm observed by
	  MAMBO2. Contours are 0.15\,MJy/sr to 2.65\,MJy/sr by
	  0.5\,MJy/sr.}
	\label{hh fig}
  \end{center}
\end{figure*}


As a complement, we also use continuum emission maps at
1.2\,mm obtained with MAMBO2, the MPIfR 117-channel
bolometer array operated at Pico Veleta
\citep{kreysa92}. These data are partially shown in a
companion paper \citep{teyssier04}, but are presented here
infull for the first time. We refer to this article for
further details about the observations. Also used in this
analysis is the ISOCAM map obtained on this source by
Abergel et al. (2002, 2003) in the LW2 filter
(5.-8.5\,$\mu$m). The corresponding maps are displayed in
Fig.~\ref{hh fig}, in combination with the H$\alpha$
coverage obtained by Reipurth \& Bally (private
communication) at the KPNO 0.9\,m telescope.

%
\section{Morphology}
\label{spatial}
%

\subsection{Spatial distribution}
\label{morpho}

Observation of an optically thin species provides a
detailed picture of the inner molecular layers of this
complex object for the first time. The spatial distribution
correlates remarkably well with the visible dust absorption
(well represented here by the H$\alpha$ image) and the
optically thin dust continuum emission at 1.2\,mm, revealing
a somewhat different shape than the Horsehead silhouette as
traced \eg\ by the BIMA CO\jone\ maps of
\citetalias{pound03} (10\arcsec$\times$8\arcsec
resolution). Also, the \cdo\ map presented here extends
further east than the \twCO\ map of these authors. In
particular, the western ridge forming the PDR appears
connected to the parental cloud through a thin
(1.5$^{\prime}$, or 0.2\,pc at a distance of 400\,pc) dust
and gas east-west filament (hereafter called the ``neck'')
exhibiting three noticeable integrated intensity peaks. This
filament penetrates further inside the L1630 cloud following
a SW-NE orientation. In this study, we distinguish between
these two sections of the neck and refer to the ``inner''
(eastmost) and ``outer'' (westmost) necks,
respectively. Similarly, we refer to the northern part of
the PDR as the ``nose'' and to its southern part as the
``mane'', following some of the naming already used by
\citetalias{pound03}. The material voids present north and
south of the outer neck will be respectively called the jaw
and the optical hole. These areas are indicated in
Fig.~\ref{position cut}. We note that our \cdo\jtwo\
integrated intensity map is much less clumpy than that in
\twCO\jone\ from \citetalias{pound03}, particularly in the
outer neck. Since \twCO\jone\ traces the lower density
material, this suggests that the external envelope is
clumpier than the inner parts evidenced by the
\cdo\jtwo. Indeed some \twCO\jone\ peaks of integrated
intensity (\eg\ in the mane) do match optical features, with
no counterpart in \cdo\jtwo. However, it is not clear
whether this is due to excitation or photo-dissociation
selective effects, or to a combination of both.

A noticeable emission peak is observed at the base of the
pillar formed by the outer neck (hereafter Peak2,
$\alpha_{2000}$=05\h41\m06\s, $\delta_{2000}=
-2$\deg27\arcmin30\arcsec), which coincides with an embedded
structure seen in absorption at 7 and 15\,$\mu$m
\citep[][see also Fig.~\ref{hh fig}]{abergel02} and a strong
emission feature in the continuum at 1.2\,mm, also prominent
in other submm continuum maps obtained by SCUBA at the JCMT
(850 and 450\,$\mu$m, Sandell private communication) and
SHARC-II at the CSO (350\,$\mu$m, Lis private
communication).  It is interesting to note that the position
of the peak associated with this clump differs in the 1.2mm
continuum and in the \cdo\jtwo\ maps, a phenomenon often
associated with molecular depletion onto grains
\citep[\eg][]{caselli99} and indicative of high densities
and cold temperatures. In the \twCO\jone\ map from
\citetalias{pound03}, an elongated maximum is seen north of
Peak2 position, but it does not coincide with either of our
\cdo\jtwo\ or continuum maxima in this area.  A detailed
study of this condensation is, however, beyond the scope of
the present paper and will be addressed in a forthcoming
work (Teyssier \& Hily-Blant, {\it in preparation}).  Three
other peaks are seen in the neck: Peak1
($\alpha_{2000}$=05\h40\m59\s, $\delta_{2000}=
-02$\deg27\arcmin18.7\arcsec), Peak3
(($\alpha_{2000}$=05\h41\m12.2\s,
$\delta_{2000}=-02$\deg25\arcmin59\arcsec), and Peak4
($\alpha_{2000}$=05\h41\m15.3\s,
$\delta_{2000}=-02$\deg25\arcmin39.5\arcsec, see also
Fig.~\ref{position cut}). We note that Peak1 has
counterparts in the 1.2mm continuum map; however, it does
not stand out at all in the interferometric map of
\citetalias{pound03}. On the other hand, Peak3 and Peak4 are
only visible in the molecular data, and in limited velocity
ranges, $[9.9:10.1]$ and $[10.3:10.6]$\,\kms\ respectively
(see Fig. \ref{chan map}).

\def\wa{0.6\textwidth}
\begin{figure*}
  \begin{center}
	\includegraphics[width=\wa,angle=270]{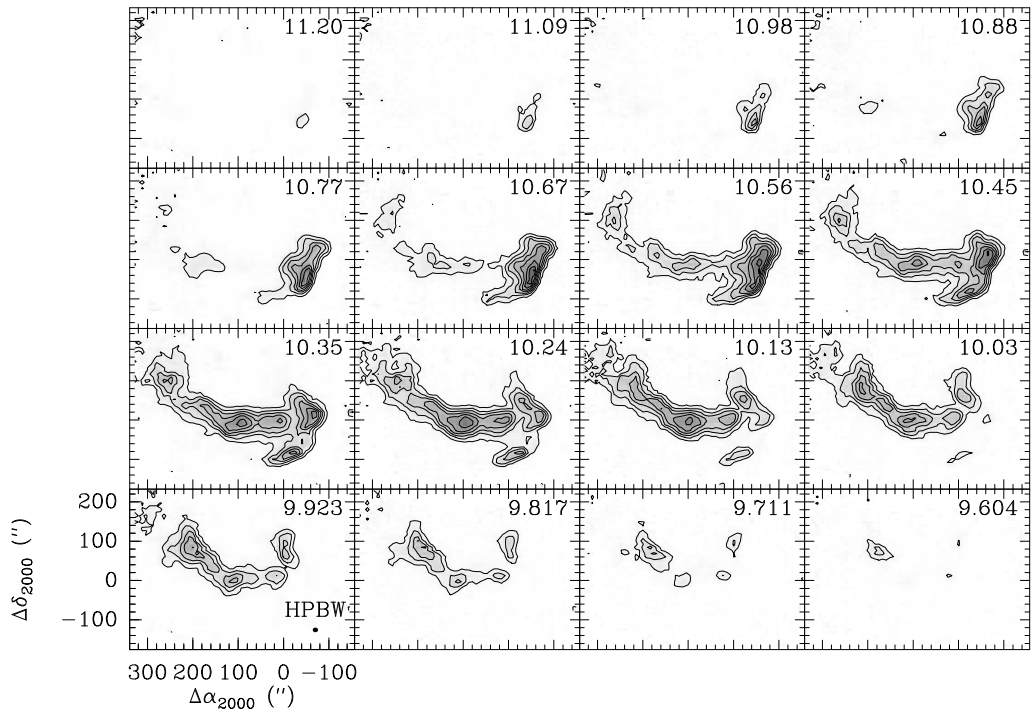}
	\caption{Channel maps of the C$^{18}$O\jtwo\ emission
	  between 9.6 and 11.2\,\kms . Contours are 1.1\,K to
	  7.4\,K by steps of 0.9\,K (\tant\ scale, rms=0.2\,K).}
	\label{chan map}
  \end{center}
\end{figure*}

\begin{figure}
  \def\wa{\hsize}
  \begin{center}
	\includegraphics[width=\wa,angle=0]{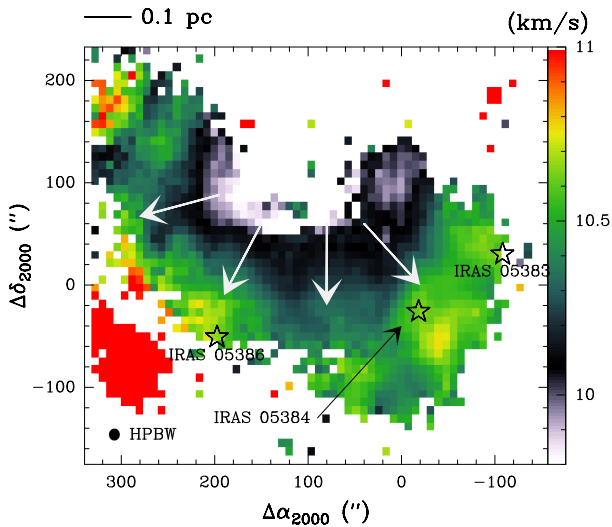}
	\caption{Map of the C$^{18}$O\jtwo\ velocity
	  centroids. Indicated by stars are three IRAS sources
	  identified by \citet{sandell86}. The arrows indicate
	  the velocity gradient orientations at different
	  positions across the neck and ridge.}
	\label{fig:centroid}
  \end{center}
\end{figure}

\begin{figure*}
  \def\wa{0.6\hsize}
  \begin{center}
	\includegraphics[width=\wa,angle=270]{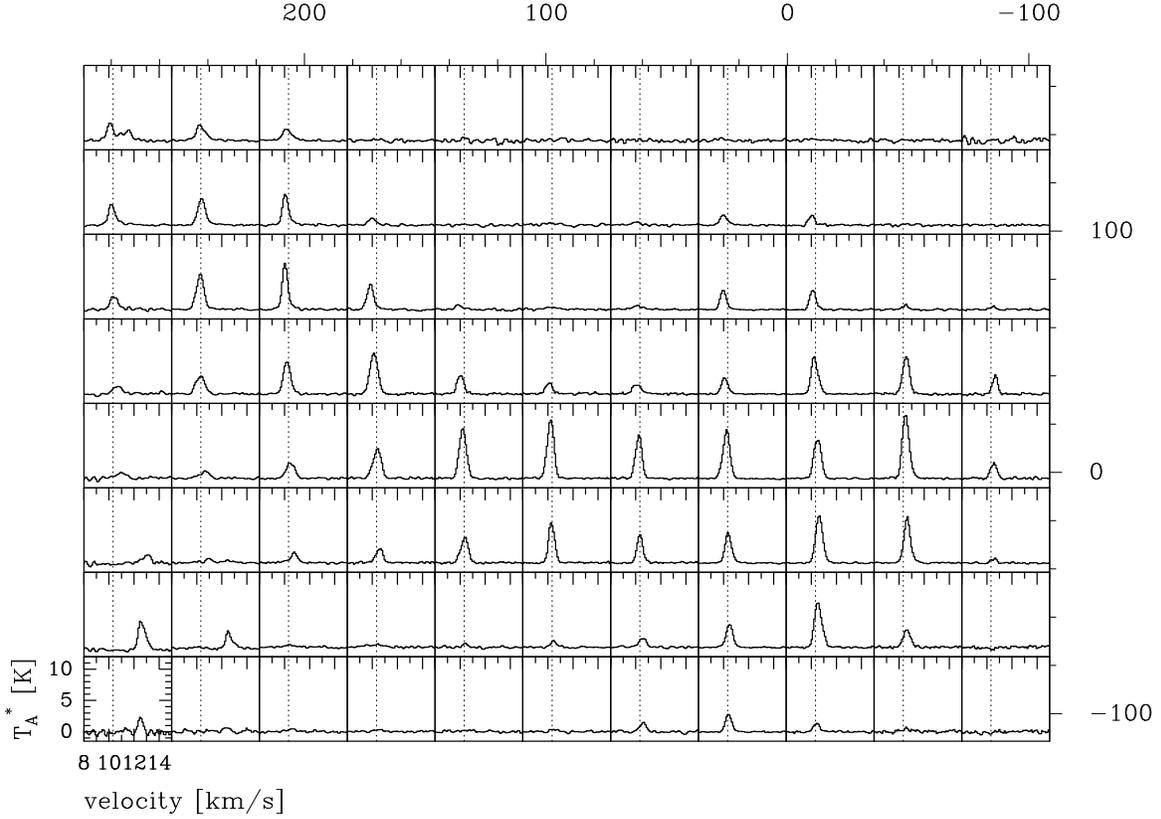}
	\caption{Map of \cdo\jtwo\ spectra averaged over areas
	  of 40\arcsec$\times$40\arcsec. The vertical dotted
	  line indicates the systemic velocity (10.33\kms).  The
	  channel width is 0.1\,\kms. Offsets are in (\arcsec).}
	\label{fig:specmean}
  \end{center}
\end{figure*}

\subsection{Physical conditions}
\label{phys condition}

This section is not meant to study the variation of the
physical parameters at small scale in detail but rather to
give an overview of the typical conditions associated to the
areas introduced above. This analysis has already been
conducted in the PDR by \cite{abergel03} and
\cite{teyssier04}, who report peak visual extinctions in the
range 12-25\,mag and densities on the order of
2\tdix{4}\ccc. However, based on high resolution \hh\ NIR
data, \citet[see also\,\citealt{habart04}]{habart03} have
shown that this last parameter could be higher by one order
of magnitude at the PDR border. At the scales of interest,
we used the 1.2mm continuum emission in order to compute the
column densities. We applied the formula presented in
\cite{motte98}, assuming dust temperatures of 35\,K in the
PDR area \citep{teyssier04} and of 10\,K for all areas
eastward of the ridge. The dust mass opacity at 1.2mm is known
to vary depending on the temperature and/or density
\citep[\eg][]{ossenkopf94,stepnik03}. To account for this
uncertainty, we considered values of this parameter in the
range 3--6\tdix{-3}\,g\,\cc. Excluding the 3 peaks
introduced above, the computed column densities vary between
some \dix{21}\,\cc\ in the most diffuse parts (distributed
over the inner and outer necks) and 1.0--2.5\tdix{22}\,\cc\
(\eg\ in the mane or the PDR). While Peak1 and Peak3 exhibit
quantities in the range of those in the PDR, the Peak2
column density is calculated to be
3.5$\pm$1.0\tdix{22}\,\cc.

In order to compare these estimates with quantities derived
from the molecular tracers, we used complementary
observations of CO\jtwo\ (G\"usten, private communication)
and \cdo\jone\ (Teyssier, private communication) at
positions of interest to run LVG simulations. Assuming
optically thick emission from CO\jone, we inferred lower
limits to the kinetic temperature to be applied to
\cdo. \tkin\ are found in the range 20-30\,K, translating
into volume densities of \dix{4}\,\ccc\ in most of the ridge
and outer neck. Somewhat lower values (factor 2-3 below) are
found in the inner neck, while Peak2 exhibits a higher
density of $n_{\rm H_2}$=4\tdix{4}\,\ccc. The \cdo\ column
density was simultaneously constrained and translated into
\hh\ column densities using the calibration described in
\cite{teyssier04}: $\Nhh = (2.5\pm0.6 \tdix{6}) \,\times\,
N({\rm C^{18}O}) +1.4\tdix{21}\,\cc$. \Nhh\ is found to be
in very good agreement with the estimates based on the dust
emission, except at the position of Peak2 where a value of
1.4\tdix{22}\,\cc\ is measured, suggestive again of
molecular depletion in the cold condensation. We estimate
from this that the quantities derived in this core from
\cdo\jtwo\ are under-estimated by a factor of at most
3. Table~\ref{tab: phys condition} compiles these values for
various areas of interest.

\begin{table}[h]
  \begin{center}
	\caption{Physical conditions in the Horsehead (see
	  Sect.~\ref{phys condition}). \newline$^{\rm (a)}$ corrected
	  from molecular depletion, $^{\rm (b)}$ from
	  \cite{abergel03}.}
	\begin{tabular}{l c c c} \hline \hline
	  \rule[-1ex]{0mm}{4ex}
	  & diffuse parts & Peak2$^{\rm (a)}$ & PDR$^{\rm (b)}$ \\ \hline
	  \rule[-2ex]{0mm}{6ex}
	  \nhh\ (\ccc) & 3--5\tdix{3} & 4\tdix{4} & 2\tdix{4} \\
	  \rule[-2ex]{0mm}{0ex}
	  \Nhh\ (\cc)  & $\approx$10$^{21}$ & $3.5\pm1.0$\tdix{22}
	  & 1.2--2.5\tdix{22}  \\\hline
	  \label{tab: phys condition}
	\end{tabular}
  \end{center}
\end{table}

We finally compared the values of \Nhh\ and n$_{\rm H_2}$ to
infer the cloud depth along the line of sight. Their ratio
indicate depths in the range 0.15--0.30\,pc, very similar to
the filament width on the sky. Within the effects of
projection, this is consistent with a more or less
cylindrical shape of the filament.

\subsection{Mass estimates}
\label{mass}

With these numbers, we estimated the overall mass of the
material traced by \cdo\jtwo. The analysis presented above
shows that $H_2$ column density estimates based on either
our 1.2mm dust continuum or the \cdo\jtwo\ integrated
intensity maps are in good agreement provided molecular
depletion onto grains is not a concern. This assumption is
fairly justified here, with the exception of the dense core
associated to Peak2. Assuming an overall conversion factor
of $\Nhh/W({\rm C^{18}O(2-1)}) = 2.5 \times
10^{21}$\,\cc/(\kkms ), the total mass derived from the
\cdo\jtwo\ integrated intensity is found to be
19\,M$_\odot$, of which 0.6\,\msol are located in the cold
core. Assuming a depletion factor on the order of 3 (see
previous section) in the Peak2 area, the total corrected
mass amounts to 20\,\msol, showing that depletion effects
are negligible as regards the total area considered. This is
lower by 25\% than the value reported by
\citetalias{pound03}, and very likely does not account for
all the material present in the nebula, since here we are
probing only the gas in the inner layers.

\begin{figure*}
  \def\wa{0.6\hsize}
  \begin{center}
	\includegraphics[width=\wa,angle=270]{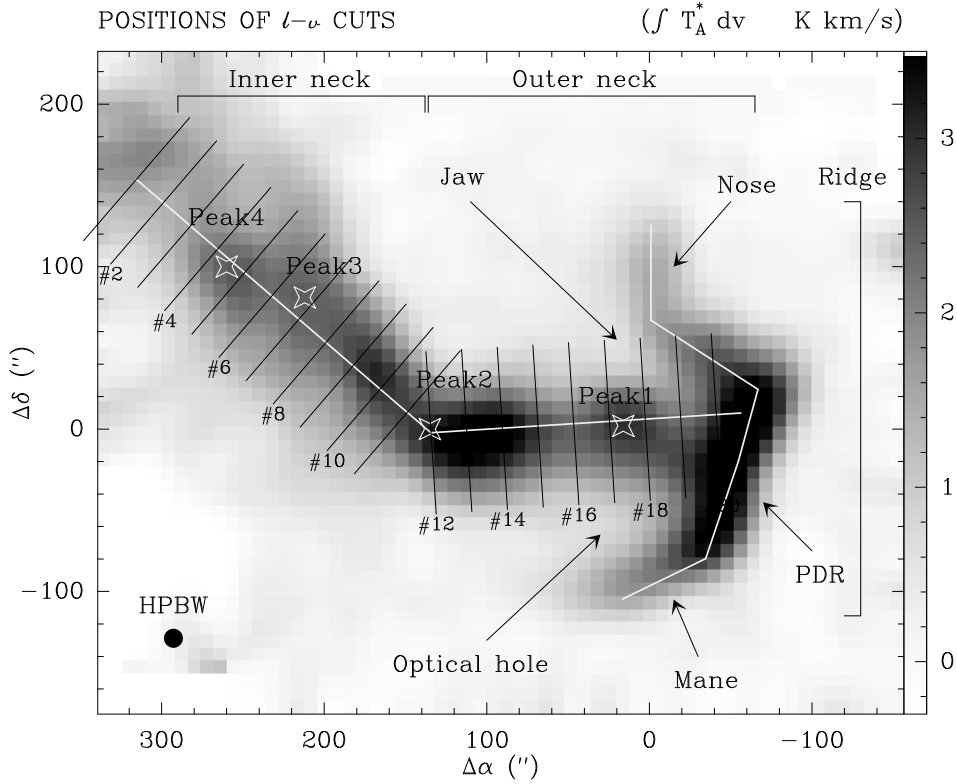}
	\caption{Position of the position-velocity cuts drawn
	  over the integrated emission map of
	  C$^{18}$O\jtwo. The full lines correspond to the cuts
	  across the neck and ridge, respectively, while the thin
	  black lines illustrate the position of the radial
	  cuts, spaced by 22\,\arcsec ($\approx2$ HPBW), with
	  some numbers referring to the cuts shown in
	  Fig.~\ref{gradv}. The crosses indicate the position of
	  the four peaks introduced in Sect.~\ref{morpho}. Also
	  indicated are the areas discussed throughout this
	  paper.}
	\label{position cut}
  \end{center}
\end{figure*}

%
\section{Study of the velocity field}
\label{velo}
%

In this section, we present the detailed analysis of the
velocity field based on the \cdo\jtwo\ data. The systemic
velocity as deduced from the integrated spectra over the
whole field is $v_0$=10.33\,\kms . The shape is very nearly
Gaussian with a total FWHM of 0.8\,\kms . This value falls
in between typical linewidth in dark clouds (nearly thermal,
0.5\,\kms ) and that of the more diffuse molecular phase
(1\,\kms ). This in fact reflects the wide range of
velocities in the nebula where the overall linewidth is
broadened by velocity gradients. Individual spectra indeed
have FWHM=0.66$\pm$0.11\kms, as illustrated in
Fig.~\ref{fig:specmean} where averaged spectra over small
areas are plotted. While the centroid velocity changes
within the nebula, the lineshape remains Gaussian.

Another picture of this effect is shown in the line centroid
map displayed in Fig.~\ref{fig:centroid}. The computation
was done in such a way that the spectral window used to
deduce the centroid is optimized for {\it each}
spectrum. This method is described in \cite{pety03}, and
consists in maximizing the signal-to-noise ratio on the
integrated area in the actual spectral window. The centroid
is then computed in the usual way, i.e. as the velocity
weighted by the temperature. As was already reported by
\citetalias{pound03}, the map in Fig.~\ref{fig:centroid}
reveals an overall North (blue-shifted emission)--South
(red-shifted emission) velocity gradient of about
4-5\,\kmspc. However, this gradient has a somewhat more
complex morphology, as its orientation seems orthogonal to a
parabolic-like curve starting (at least within our map
boundaries) from the inner neck basis and slowly folding
towards the Horse's nose. The gradient orientation is thus
swivelling east-west, as indicated by the arrows drawn on
Fig.~\ref{fig:centroid}. The NE-SW gradient mentioned by
\citetalias{pound03}\ thus only applies to the western part
of the dark protrusion. The data also indicate that both red
and blue velocity components seem to be wrapped around an
axis coincident with the neck shape. It is interesting to
note the coincidence between the peaks in velocity centroids
(around 10.8\,\kms ) and the locations of three of the IRAS
stars identified in the field (Fig.~\ref{fig:centroid}). A
similar phenomenon could already be seen on the map of
\citetalias{pound03}. We do not, however, see any obvious
explanation for this effect.

The channel map displayed in Fig.~\ref{chan map} shows the
structures associated to the \cdo\jtwo\ velocity field.  The
ridge and the neck appear clumpy, with the intensity peaks
moving when going from one velocity channel to the
other. This is indicative of a rich velocity field inside
the structure.
In order to identify the mechanisms associated to this gas
motion, we study position-velocity maps performed along
various dedicated cuts in the following sections.

\subsection{Longitudinal cuts}
\label{longitudinal cuts}

We first consider cuts performed along the filamentary
structures introduced in Sect.~\ref{morpho}. The positions
of these two cuts (respectively along the neck and the
western ridge) are displayed in Fig.~\ref{position
cut}. Although there is some subjectivity in the choice of
these segments, we followed, as much as possible, the
morphology traced by the filamentary structures in the
integrated intensity map.

\begin{figure}
  \def\wa{0.9\hsize}
  \begin{center}
	\includegraphics[width=\wa,angle=0]{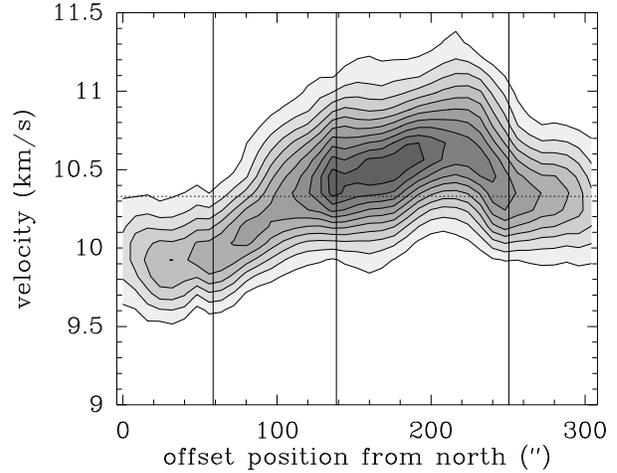}
	\caption{\cdo\jtwo\ position-velocity cut performed
	  along the western ridge and the mane (from north to
	  south, Fig.~\ref{position cut}). The dashed horizontal
	  line indicates the systemic velocity $v_0$. The three
	  vertical lines delineate the four segments forming the
	  cut. A large (+2.6\,\kmspc) north--south velocity
	  gradient is observed until the mane, where the
	  gradient is suddenly inverted to finally end up at the
	  systemic velocity with a slope of
	  --4.3\,\kmspc. Contours start at 3$\sigma$\,(0.6\,K),
	  in steps of 3$\sigma$.}
	\label{lv ridge}
  \end{center}
\end{figure}

Figure~\ref{lv ridge} displays the first of these cuts
computed along the western ridge, showing a large
north-south velocity gradient extending over more than
0.4\,pc. The relatively linear slope of this gradient (on
the order of +2.6\,\kmspc) suggests that the material is
rotating.  The PDR and nose thus seem to be wrapped around
an axis coincident with the outer neck. Further along the
cut, the velocities associated to the gas traced in the mane
experience a fast drop with a gradient of
$-4.3$\,\kmspc. The mane thus appears to be braked back to
the systemic velocity found in the filament. We discuss in
Sect.~\ref{discussion} an interpretation of this peculiar
behaviour.

\begin{figure}
  \def\wa{0.9\hsize}
  \begin{center}
	\includegraphics[width=\wa,angle=0]{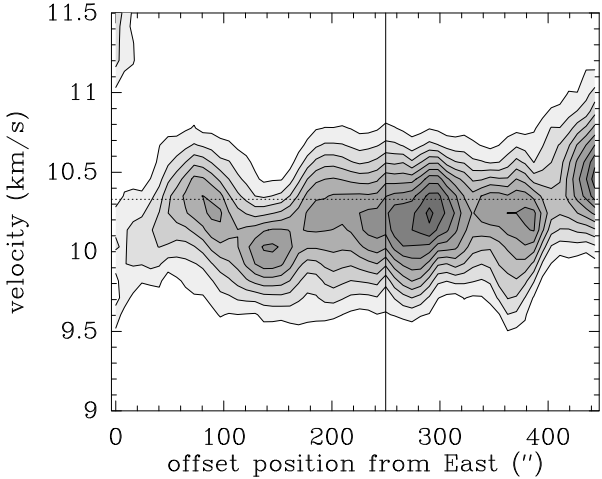}
	\caption{Same as Fig.~\ref{lv ridge} along the entire
	  neck (from east to west, Fig.~\ref{position cut}). The
	  dashed line indicates the systemic velocity, while the
	  vertical one shows the position of Peak2 (see
	  Sect.~\ref{morpho}), where the longitudinal cut
	  changes direction.  Positions of peaks 4 to 1 are at
	  offsets 80, 140, 250 and 380\,\arcsec\
	  respectively. Recall that Peak2 is not a maximum in
	  the \cdo\jtwo\ integrated intensity map and was
	  inferred from the continuum data. Note the 3 different
	  velocity gradients (positive, negative, positive) as
	  one moves from 0 to 200\,\arcsec (see text for
	  details). Contours start at 3$\sigma$\,(0.6\,K), in
	  steps of 3$\sigma$.}
	\label{lv neck}
  \end{center}
\end{figure}

The respective $l-v$ map in the neck is shown in
Fig.~\ref{lv neck}. In the gas associated to the outer neck,
departures from the systemic velocity are less pronounced
than in the ridge. However a close look at the inner neck
indicates that the velocity gradient appears to change sign
regularly (amplitude of approximately 2.5\,\kmspc, rough
period of 2.5\arcmin, or 0.3\,pc). The nodes of this
modulation are found at a common velocity of 10.2\,\kms (see
Fig.~\ref{lv neck}). It is interesting to note that
Peaks~3--4 are located where the velocity gradient changes
sign. This gradient then experiences another rapid flip as
one penetrates the L1630 cloud (new value of --4.8\,\kmspc)
further.


\subsection{Radial cuts}
\label{radial cuts}

\begin{figure}
  \def\wa{1.\hsize}
  \begin{center}
	\includegraphics[width=\wa,angle=0]{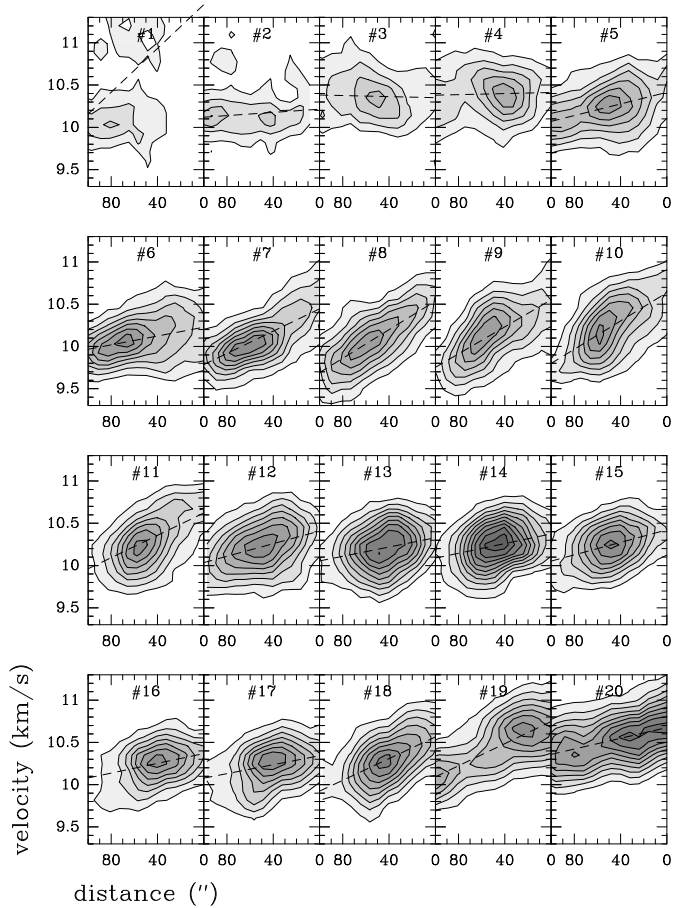}
	\caption{Transverse \cdo\jtwo\ position-velocity cuts
	  performed across the neck, corresponding to the black
	  thin lines of Fig.~\ref{position cut}. The cuts are
	  separated by 22\arcsec. The x-axis offsets increase
	  northwise, so that the velocity decreases from South
	  to North. Dashed lines indicate the results of the
	  bi-dimensional fits to the $l-v$ diagram (see text for
	  details), showing how the radial velocity gradient
	  evolves as one moves from East to West (cf
	  Fig.~\ref{gradv}). Contours start at 3$\sigma$
	  (0.6\,K), in steps of 3$\sigma$. }
	\label{lv radial}
  \end{center}
\end{figure}

Radial cuts across the neck allow a probe of the gas motion
around the axis defined by the filament projected shape. To
this aim, we have constructed $l-v$ maps on a series of
small slits perpendicular to the neck longitudinal cut
analyzed previously. Their positions and associated numbers
are indicated in Fig.~\ref{position cut}. The corresponding
collection of maps is gathered in Fig.~\ref{lv radial}. In
most cases, the velocity scales almost linearly with the
radial position, again indicative of rigid body rotation and
confirming that the neck itself is spinning around its own
axis. However, some cuts (\eg\ \#6, \#7, \#11, \#12) depart
from that linear behaviour, suggesting that differential
rotation may affect the transverse gas motion in these
areas. Such velocity gradients across the neck could also
result from a velocity shear perpendicular to the neck. This
interpretation will be compared to rotation in
Sect.\ref{shear}.

It is also interesting to note that the velocity gradient
observed at each position is varying when moving from east
to west. In order to quantify this variation, and motivated
by the assumption of rigid body rotation, we performed a
systematic linear fit of each individual $l-v$ cut and
interpreted this slope of as an angular velocity.  In each
cut, spectral lines were fitted by Gaussians, and their
central velocity was stored.  This set of central velocities
was then weighted taking into account both the
signal-to-noise ratio (SNR) of the data and the error on the
centroid itself, and finally fitted by a linear slope. The
result of this computation is illustrated in
Fig.~\ref{gradv}, where gradients are plotted against the
distance along the neck. The resulting plot reveals a
striking shape: while the outer neck appears in rigid body
rotation at 1.5\,\kmspc\ (corresponding to a rotation period
of 4\,Myrs, roughly equal to the survival time of the
Horsehead, 5\,Myrs, estimated by \citetalias{pound03}),
parts of the inner neck experience a sharp increase in
velocity gradient when moving from east to west (from 1.5 to
4\,\kmspc\ in about than 0.1\,pc). This rapid jump is
followed by a similar decrease back to the 1.5\,\kmspc\
velocity gradient. Beyond this point, the radial cuts
associated to the westmost part of the inner neck are found
to have a 0 velocity gradient (within the slope fit
uncertainty). It is to be noted that all three positions of
Peaks 1, 2, and 4 are found at the common velocity gradient
of 1.5\,\kmspc\ (it is less true for Peak1, though), while
Peak3 is found at the high velocity gradient of
4\,\kmspc. Interestingly enough, Peak3 sits almost in the
middle of the roughly symmetric gradient increase bracketed
by Peaks 2 and 4. At the eastmost offsets, the radial cuts
exhibit another gradient increase, but this one is very
likely due to the known large gradient found along the mane
and nose, and is less meaningful in terms of radial gradient
across the neck.

It is also interesting to compare the ratio of the
gravitational to centrifugal forces in various areas of the
Horsehead. This ratio is given by $\zeta=\pi G
\langle\rho\rangle/\omega^2$ (with $\langle\rho\rangle$ the
mean mass density and $\omega$ the angular velocity),
translating into $\zeta\approx 10$ in the outer neck, while
$\zeta\approx 1$ near Peak2. This means that gravity is able
to ensure the confinement in the outer neck, while it is only
marginally sufficient near Peak2.

%
\section{Discussion}
\label{discussion}
%

\subsection{Formation and evolution of the Horsehead nebula}
\label{genesis}
  In the scenario proposed by \cite{reipurth84}, the
  Horsehead is assimilated to an early Bok globule emerging
  from its parental cloud via the eroding incident radiation
  field emitted by $\sigma$Ori. These authors also propose
  that the jaw cavity was the result of a collimated
  outflow, but this seems now highly unlikely \citep[\eg\
  ][]{warren85}. Based on their CO\jone\ data,
  \citetalias{pound03} discuss the possible origin of
  this dark cloud, and conclude that both the work of
  instabilities and of the ionization front could be
  considered. They propose that the initial density
  enhancement in the parental cloud was asymmetric, at least
  at the western end of the neck, and that this material has
  been pushed out on either side of the neck, resulting in
  the mane and the jaw. They also deduce that the
  Horsehead's neck is not axisymmetric but rather presents
  an oval cross-section.
  Their conclusions are, however, related to the general
  shape of the cloud, so they may have missed some of the
  details in the inner layers of the cloud due to the high
  opacity of their tracer. The data obtained in \cdo\ thus
  shed new light on the formation and evolution of the
  Horsehead.

  Based on the velocity cuts described in the previous
  sections, the overall picture is that of an elongated
  structure (the neck) spinning around its own axis, and
  connected at its western end to a ridge (mane to nose) in
  rotation around this same axis. We found that this
  filament is roughly axisymmetric with a projected
  diameter $D\approx0.15$\,pc and a depth
  $L\approx0.15-0.30$\,pc.  In the \citetalias{pound03}\
  scenario, all that is needed is some asymmetry at the
  front end of the neck, therefore it is not inconsistent
  with the present conclusion.

  The reason the Horsehead followed such a peculiar route
  from L1630 might be linked to a pre-existing rotating
  velocity field in the region of the parental cloud that
  resisted the incident radiation field and gave birth to
  the protrusion, a possibility also mentioned by
  \citetalias{pound03}. As rotation proceeded, the
  centrifugal force has progressively detached the nose and
  the mane from the neck. This scenario requires initial
  density inhomogeneities within the pillar: the densest
  parts have formed the neck, mane, and nose, while the most
  tenuous ones remained as the jaw cavity and the optical
  hole. We note that this inhomogeneity is also required by
  the scenario reported by \citetalias{pound03}. This
  picture is moreover consistent with the gradient seen in
  the longitudinal $l-v$ cut shown in Fig.~\ref{lv ridge}:
  the material extending from the nose down to the middle of
  the PDR (offsets 0 to 200\arcsec) is rotating, with the
  nose blue-shifted and part of the mane
  red-shifted. The southern end of the mane (offsets 200 to
  300 \arcsec\ in Fig~\ref{lv ridge}) is, however, braked back
  to the neck velocity. This suggests that part of the mane
  has remained attached to the pillar (their physical link
  is already obvious from various tracers, see Fig.~\ref{hh
  fig}), resulting in the material loop seen \eg\ in the
  optical. As a consequence of this mechanism, the nose and
  the mane cannot lie in the plane of the sky as the neck
  presumably does. Rather, the nose would be in the
  foreground, while part of the mane would be in the
  background. The fact that the nose appears totally
  detached from the neck also indicates that the material
  void associated to the actual jaw was more pronounced than
  that in the optical hole. Obviously, the continued eroding
  work of $\sigma$Ori has proceeded during this process. It
  must thus have also contributed to the shaping of the
  structure we see in projection on the plane of the
  sky. This must be particularly true in the jaw and the
  optical hole, where the initial density contrast within
  the pristine pillar may thus have been enhanced.

  In complement to this effect, the nose may also be pushed
  towards us by the rocket effect resulting from eventual
  background ionization. However, the amplitude of this
  effect must be negligible with respect to the prevailing
  velocity field since part of the mane remains
  red-shifted. Moreover, the rocket effect would also affect
  the neck which is seen to be rotating.  We therefore
  discard the possibility of a non-negligible background
  ionization flux.

\subsection{Rotation \vs\ shear}
\label{shear}

We discuss here the question whether the velocity gradient
revealed in the neck could be the signature of a velocity
shear in a direction perpendicular to both the neck and the
line of sight, rather than a rotation of the filamentary
neck. We see at least three arguments favouring the rotation
interpretation.

First of all, we have proposed in the previous section that
both the (likely combined) effect of an abrasive ionizing
front and that of a pre-existing velocity field may have
contributed to the peculiar shaping of the Horsehead
nebula. It is very unlikely that the ionization front
arising from $\sigma$Ori could have generated a shear
perpendicular to the line-of-sight, so that a velocity shear
present in the neck should have its origin in the parental
cloud. A careful look at the position-velocity cuts of
\cite{kramer96} in \thCO\jtwo\ show no velocity gradient in
the vicinity of the Horsehead nebula (Right Ascension
offsets between -4 and -12\,arcmin). There would thus be no
shear in the parental cloud, and these would likely discard
its existence in the neck.

A second argument consists in considering the morphological
consequence of such a shear. Indeed, after some time, a
shear in the neck would result in a filament far from being
cylindrical, which is in disagreement with the result
inferred from the column density calculation. The timescale
for this motion can be roughly estimated from that of
momentum transfer between the ionized and the condensed and
cold gas. Assuming an average collisional cross-section for
H--H$^+$ encounters, $\langle
u\sigma\rangle=2.2\tdix{-9}\ccc\,$s \citep{spitzer1978}, and
adopting a mean density of proton $n_f$ as found in the
Horsehead, one yields $\Delta t \approx 2/(n_f\langle
u\sigma\rangle)\approx \dix{6}$s, which even if
underestimated by orders of magnitude, would be  short
enough by far to have disrupted the cylinder shape one still sees
today.

Finally, if despite the arguments given above, shear would
still play a role in the dynamic at work in the neck, and
the torque it would exert on the mane and nose would in any
case result in rotation in the filament.

\begin{figure}
  \def\wa{\hsize}
  \begin{center}
	\includegraphics[width=\wa,angle=0]{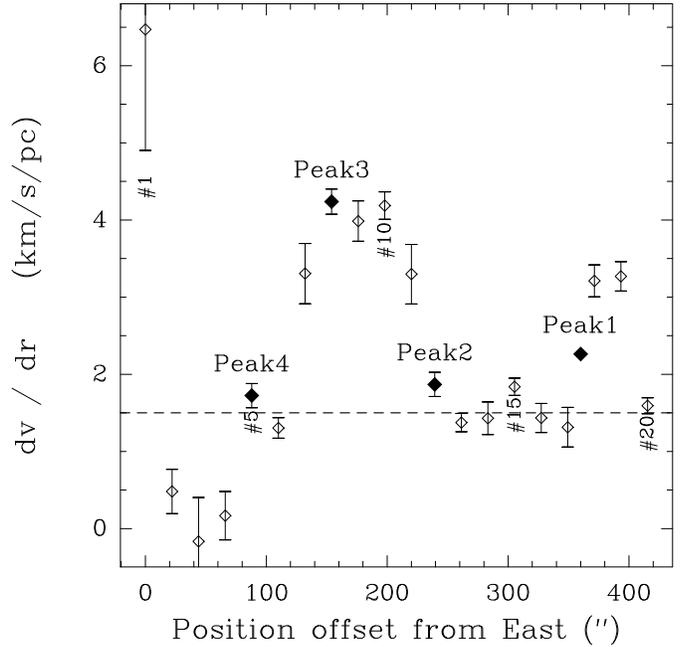}
	\caption{Evolution of the transverse velocity gradient
	  with the distance along the neck. The gradients and
	  associated error bars were computed following the
	  method described in Sect.~\ref{radial cuts}. The
	  dashed line shows the constant gradient of
	  +1.5\,\kmspc, indicating the area suggestive of rigid
	  body rotation. Also shown are the peaks discussed in
	  this paper. Some cut numbers are recalled for the
	  purpose of comparison.}
	\label{gradv}
  \end{center}
\end{figure}

\subsection{Dense core formation in the Horsehead}
\label{dense core}

  Dense cores are routinely observed in filaments; however,
  what triggers their formation is still an open
  issue. Fragmentation of gaseous cylinders resulting from
  periodic instabilities is often invoked, but their nature
  is uncertain, be they gravitationally or magnetically
  dominated, or even a combination of both. Determining the
  balance between these two forces appears to be a critical
  issue.

  We have shown that the Horsehead neck is a filament
  harboring (at least) four density peaks (as evidenced by
  both 1.2mm and \cdo\ maps). The longitudinal cuts in the
  neck (Fig.~\ref{lv neck}) show that the longitudinal
  velocity is not uniform but instead varies periodically in
  the inner neck (period\,$\approx 0.3$\,pc), around
  $V=10.2$\,\kms\ close to the systemic velocity. This
  indicates that the filament is not located strictly in the
  plane of the sky. This oscillation could be explained
  considering matter that flows along a filament regularly
  bent with respect to the plane of the sky. Moreover, it is
  worth mentioning the presence of line wings around Peak3
  (see offsets (200\arcsec,80\arcsec) in
  Fig.~\ref{fig:specmean}). However, the spatial resolution
  of our data is not sufficient to conclude a scenario such
  as outflow or inflow.

  The radial cuts shown in Fig.~\ref{lv radial} indicate
  that the phenomena at work are more complex: the
  longitudinal gradients are combined to rotation. This
  rotation is mainly that of a rigid body, although the
  angular velocity experiences a sharp increase from 1.5 to
  4\,\kmspc\ (Fig.~\ref{lv radial} and
  Fig.~\ref{gradv}). Interestingly, the longitudinal
  gradient appears to change signs at the positions of Peak3
  and 4 (Fig.~\ref{lv neck}), positions that also appear to
  play a particular role in the picture of radial gradients
  (Fig.~\ref{gradv}). It suggests a dynamical link between
  the formation of the condensations observed in the neck
  and the material flow revealed by the complex velocity
  structure of the filament. Finally, we note that another
  likely effect of the combination of both longitudinal and
  radial motions is the morphology change observed between,
  \eg, cuts \#8 and \#14 (Fig.~\ref{lv radial}): cut \#14 is
  much more rounded than the straight contours of cut
  \#8. At this stage it becomes difficult to derive an
  accurate picture accounting for the combination of all
  these phenomena. We believe that detailed interpretation
  of all these signatures requires dedicated numerical
  modelling.

  We can, however, already infer preliminary constraints about
  some of the principal ingredients of this modelling,
  namely gravity and magnetic field. In cylindrical
  geometry, the virial mass per unit length may be written
  \citep[\eg][]{fiege00a}:
  \begin{equation}
	m_{\rm vir} = \frac{2\langle\sigma^2\rangle}{G}.
  \end{equation}
  Assuming $\langle\sigma^2\rangle\approx\sigma_v^2$, this
  gives
  \begin{equation}
	m_{\rm vir} = 4.6 \left(\frac{\sigma_v}{0.1\,\kms}\right)^2
	\,\,\msol/pc.
  \end{equation}
  For the velocity dispersion, we apply the \citet{fuller92}
  formula:
  \begin{equation}
	\sigma_v^2 = \sigma_{v,\rm obs}^2 + k_B\tkin
	\left(\frac{1}{\tilde m}-\frac{1}{m_{\rm obs}} \right)
  \end{equation}
  where ${\tilde m}$ is the mean mass of molecular gas,
  taken here equal to 2.33; $m_{\rm obs}$ is the mass of the
  molecular tracer namely \cdo; and \tkin\ the kinetic
  temperature. We adopt \tkin\,=\,20K and $\sigma_{v,\rm
  obs}=0.66/2.35$\,\kms, resulting in a virial mass $m_{\rm
  vir}\approx 65\msol/$pc. On the other hand, the mass per
  unit length we can deduce from the density and size (see
  Sect. \ref{phys condition}) is
  \begin{equation}
	m = 18 \,\,%
	\left(\frac{\nh}{\dix{4}\,\ccc}\right)%
	\left(\frac{R}{0.1\,pc}\right)%
	\,\,\msol/pc.
  \end{equation}
  Taking \nhh~=~5\tdix{3}\,\ccc\ and $R \approx 0.2$\,pc
  results in mass per unit length
  $m\approx$~20~\msol/pc. The filament is not
  self-gravitating, since $m/m_{\rm vir} \approx 0.3$. This
  value falls in the range of ratio presented by
  \citet{fiege00a}. This again raises the question of the
  confinement of the filament. However, the ratio is not
  that small, and the gravity is expected to play a role in
  the dynamics of the filament.

  To our knowledge, no estimation of the magnetic field
  strength is available for the Horsehead. Polarized
  absorbed starlight is reported in \cite{warren85},
  completed by the large-scale dataset of
  \cite{zaritsky87}. In the Horsehead, the polarization is
  probably due only to alignment of absorbing dust
  grains. The transmitted light is therefore aligned with
  the magnetic fields projected on the plane of the
  sky. Obviously, this technique does not allow detection of
  any polarization in the densest parts of the
  Horsehead. The overall picture is that of a
  $\vec{B}$-field oriented nearly North-South \ie\
  perpendicular to the filament. Moreover, it coincides with
  the large-scale field. In the nose area, however, the
  polarization vectors appear perpendicular to the structure
  with an angle $\approx45^{\circ}$ with respect to the
  large-scale field. The nose may thus have bent the field
  lines in its centrifugal motion.  In the filament, though,
  we cannot yet distinguish between a toroidal component
  threading the neck and pure transverse magnetic field
  across the neck itself. It is worth noting that the same
  is observed in some dark clouds in the Taurus and in
  Ophiuchus: the magnetic fields orientation at the scale of
  the dark clouds fits into smooth, larger-scale fields. In
  the Taurus, the magnetic fields appear perpendicular to
  the cloud's long axis, while in Ophiuchus it would be
  parallel \citep{heiles92}.

  We therefore think that both gravity and magnetic fields
  are expected to play a role in the confinement and dense
  core formation in the Horsehead nebula.

%
\section{Conclusions}
\label{conclusion}
%

We have studied the morphology and velocity structure of the
Horsehead nebula using new observations at high frequency
and spatial resolution in the \jtwo\ transition of \cdo. Our
conclusions can be summarized as follows:
\begin{enumerate}
\item the emission of the likely optically thin tracer
reveals a filamentary morphology similar to the one seen in
dust millimetre continuum emission and visual
extinction. The cloud appears as a curved ridge (the PDR)
connected to L1630 by a thin (0.15\,pc diameter) pillar
hosting several condensations. One of them (Peak2) also
appears in absorption in the mid-IR emission at 7\,$\mu$m.
\item we note that our thin \cdo\jtwo\ map is much less
 clumpy than the possibly thick \twCO\jone\ from
 \citetalias{pound03}, a phenomenon that could be due to
 either excitation or photodissociation selective effects or
 the combination of both.
\item these cores are density peaks ($n_{\rm H_2}$ in the
range $2-4\tdix{4}$\,\ccc) embedded in a medium that is more
diffuse by a factor of 2-3 on average. Kinetic temperatures
in these cores are found in the range 20-30\,K, although the
brightest of them (Peak2) shows hints of \cdo~depletion onto
grains, pointing towards a somewhat colder environment.
\item the overall shape of the filament is found to be
fairly cylindrical with a diameter of about
0.15--0.30\,pc. The total mass of the \cdo\ gas probed here
is 20\,\msol, of which the largest fraction was found to be
in the gaseous phase.
\item the centroid velocity map suggests that the gas in the
  Horsehead is wrapped around an axis coincident with the
  pillar. The cloud also presents various velocity gradients
  along the filaments described above. While the western
  ridge presents two strong north-south gradients inverting
  sign at the rise of the mane, the east-west pillar
  exhibits a more complex shape where gradient oscillations
  occur on a regular basis and seem bracketed by the
  location of the embedded condensations.
\item tranverse gradients are prominent almost everywhere
  across the pillar, but some of them depart from linear
  shape and might indicate differential rotation. The
  amplitude of the gradient is on average constant
  (1.5\,\kmspc, corresponding to a rotation period of 4
  Myrs, comparable to the 5\,Myrs lifetime estimated by
  \citetalias{pound03}) at either end of the neck, but a
  sharp increase to 4\,\kmspc\ is observed between two of
  the embedded peaks. Reasons for this peculiar behaviour
  are still unclear.
\item this complex dynamic picture raises the question of
  whether the origin of this protrusion could be linked to a
  pre-existent velocity field that progressively separated
  the mane and nose from the neck via centrifugal
  effects. The presence of several condensations unveiled
  inside the filament also supports the idea that dense
  cores are forming in the neck, and their characteristics
  share several similarities with modelling works on that
  topic. Their origin is very likely due to the combination
  of longitudinal (infall) and radial motions, but the
  actual phenomena at work may even be more complex, as the
  magnetic field is expected to also play a role in this
  process. Moreover, the origin of the initial velocity
  field invoked in this work remained unknown.
\end{enumerate}

Firmer conclusions on the formation scenario of such a
protrusion should, however, await comparison with numerical
models, and the relatively simple geometry of the Horsehead
neck should be a particularly appealing application.

\begin{acknowledgements}
  We thank the referee, Marc Pound, for constructive remarks
  that allowed us to improve and clarify several points in
  this paper. We would like to thank the HERA team for
  making the collection of the data used in this study
  possible. We also thank P.~Hennebelle for constructive
  discussion of hydrodynamic models for dense core
  formation, as well as A.~Abergel for providing us with
  part of the data presented here. We also thank B.~Reipurth
  and J.~Bally for making their H$\alpha$ map available.
\end{acknowledgements}

\bibliographystyle{aa}
\bibliography{hh_vel}

\begin{thebibliography}{35}
\expandafter\ifx\csname natexlab\endcsname\relax\def\natexlab#1{#1}\fi

\bibitem[{{Abergel} {et~al.}(2002){Abergel}, {Bernard}, {Boulanger},
  {Cesarsky}, {Falgarone}, {Jones}, {Miville-Deschenes}, {Perault}, {Puget},
  {Huldtgren}, {Kaas}, {Nordh}, {Olofsson}, {Andr{\' e}}, {Bontemps}, {Casali},
  {Cesarsky}, {Copet}, {Davies}, {Montmerle}, {Persi}, \&
  {Sibille}}]{abergel02}
{Abergel}, A., {Bernard}, J.~P., {Boulanger}, F., {et~al.} 2002, \aap, 389, 239

\bibitem[{{Abergel} {et~al.}(2003){Abergel}, {Teyssier}, {Bernard},
  {Boulanger}, {Coulais}, {Fosse}, {Falgarone}, {Gerin}, {Perault}, {Puget},
  {Nordh}, {Olofsson}, {Huldtgren}, {Kaas}, {Andr{\' e}}, {Bontemps}, {Casali},
  {Cesarsky}, {Copet}, {Davies}, {Montmerle}, {Persi}, \&
  {Sibille}}]{abergel03}
{Abergel}, A., {Teyssier}, D., {Bernard}, J.~P., {et~al.} 2003, \aap, 410, 577

\bibitem[{{Caselli} {et~al.}(1999){Caselli}, {Walmsley}, {Tafalla}, {Dore}, \&
  {Myers}}]{caselli99}
{Caselli}, P., {Walmsley}, C.~M., {Tafalla}, M., {Dore}, L., \& {Myers}, P.~C.
  1999, \apjl, 523, L165

\bibitem[{{Chini} {et~al.}(1997){Chini}, {Reipurth}, {Ward-Thompson}, {Bally},
  {Nyman}, {Sievers}, \& {Billawala}}]{chini97}
{Chini}, R., {Reipurth}, B., {Ward-Thompson}, D., {et~al.} 1997, \apjl, 474,
  L135

\bibitem[{{Fiege} \& {Pudritz}(2000)}]{fiege00a}
{Fiege}, J.~D. \& {Pudritz}, R.~E. 2000, \mnras, 311, 85

\bibitem[{{Fuller} \& {Myers}(1992)}]{fuller92}
{Fuller}, G.~A. \& {Myers}, P.~C. 1992, \apj, 384, 523

\bibitem[{{Habart} {et~al.}(2004){Habart}, {Abergel}, {Teyssier}, \&
  {Walmsley}}]{habart04}
{Habart}, E., {Abergel}, A., {Teyssier}, D., \& {Walmsley}. 2004, \aap, in
  press

\bibitem[{{Habart} {et~al.}(2003){Habart}, {Abergel}, {Walmsley}, {Teyssier},
  \& {Pety}}]{habart03}
{Habart}, E., {Abergel}, A., {Walmsley}, C., {Teyssier}, D., \& {Pety}, J.
  2003, in The dense interstellar medium in galaxies, 443--446

\bibitem[{{Heiles} {et~al.}(1992){Heiles}, {Goodman}, {McKee}, \&
  G.}]{heiles92}
{Heiles}, C., {Goodman}, A.~A., {McKee}, C.~F., \& G., Z.~E. 1992, in
  Protostars and Planets III, 279

\bibitem[{{Hennebelle}(2003)}]{hennebelle03}
{Hennebelle}, P. 2003, \aap, 397, 381

\bibitem[{{Kramer} {et~al.}(1996){Kramer}, {Stutzki}, \&
  {Winnewisser}}]{kramer96}
{Kramer}, C., {Stutzki}, J., \& {Winnewisser}, G. 1996, \aap, 307, 915

\bibitem[{{Kreysa}(1992)}]{kreysa92}
{Kreysa}, E. 1992, in ESA SP-356: Photon Detectors for Space Instrumentation,
  207--210

\bibitem[{{Loren}(1989a)}]{loren89a}
{Loren}, R.~B. 1989a, \apj, 338, 902

\bibitem[{{Loren}(1989b)}]{loren89b}
{Loren}, R.~B. 1989b, \apj, 338, 925

\bibitem[{{Motte} {et~al.}(1998){Motte}, {Andre}, \& {Neri}}]{motte98}
{Motte}, F., {Andre}, P., \& {Neri}, R. 1998, \aap, 336, 150

\bibitem[{{Nakamura} {et~al.}(1993){Nakamura}, {Hanawa}, \&
  {Nakano}}]{nakamura93}
{Nakamura}, F., {Hanawa}, T., \& {Nakano}, T. 1993, \pasj, 45, 551

\bibitem[{{Obayashi} {et~al.}(1998){Obayashi}, {Fukui}, {Kun}, {Sato}, \&
  {Yonekura}}]{obayashi98}
{Obayashi}, A., {Fukui}, Y., {Kun}, M., {Sato}, F., \& {Yonekura}, Y. 1998,
  \aj, 115, 274

\bibitem[{{Onishi} {et~al.}(1999){Onishi}, {Mizuno}, \& {Fukui}}]{onishi99}
{Onishi}, T., {Mizuno}, A., \& {Fukui}, Y. 1999, \pasj, 51, 257

\bibitem[{{Onishi} {et~al.}(1998){Onishi}, {Mizuno}, {Kawamura}, {Ogawa}, \&
  {Fukui}}]{onishi98}
{Onishi}, T., {Mizuno}, A., {Kawamura}, A., {Ogawa}, H., \& {Fukui}, Y. 1998,
  \apj, 502, 296

\bibitem[{{Ossenkopf} \& {Henning}(1994)}]{ossenkopf94}
{Ossenkopf}, V. \& {Henning}, T. 1994, \aap, 291, 943

\bibitem[{{Ostriker}(1964)}]{ostriker64}
{Ostriker}, J. 1964, \apj, 140, 1056

\bibitem[{{Pety} \& {Falgarone}(2003)}]{pety03}
{Pety}, J. \& {Falgarone}, E. 2003, \aap, 412, 417

\bibitem[{{Pound} {et~al.}(2003){Pound}, {Reipurth}, \& {Bally}}]{pound03}
{Pound}, M.~W., {Reipurth}, B., \& {Bally}, J. 2003, \aj, 125, 2108

\bibitem[{{Reipurth} \& {Bouchet}(1984)}]{reipurth84}
{Reipurth}, B. \& {Bouchet}, P. 1984, \aap, 137, L1

\bibitem[{{Sandell} {et~al.}(1986){Sandell}, {Reipurth}, {Menten}, {Walmsley},
  \& {Ungerechts}}]{sandell86}
{Sandell}, G., {Reipurth}, B., {Menten}, C., {Walmsley}, M., \& {Ungerechts},
  H. 1986, in ASSL Vol. 124: Light on Dark Matter, 295

\bibitem[{{Schuster} {et~al.}(2003){Schuster}, {Greve}, {Hily-Blant},
  {Planesas}, {Sievers}, {Thum}, \& {Wiesemeyer}}]{schuster03}
{Schuster}, K., {Greve}, A., {Hily-Blant}, P., {et~al.} 2003, IRAM technical
  report

\bibitem[{{Schuster} {et~al.}(2004){Schuster}, {Boucher}, {Brunswig}, {Carter},
  {Chenu}, {Foullieux}, {Greve}, {John}, {Lazareff}, {Navarro}, {Perrigouard},
  {Pollet}, {Sievers}, {Thum}, \& {Wiesemeyer}}]{schuster04}
{Schuster}, K.-F., {Boucher}, C., {Brunswig}, W., {et~al.} 2004, \aap, 423,
  1171

\bibitem[{{Spitzer}(1978)}]{spitzer1978}
{Spitzer}, L. 1978, {Physical processes in the interstellar medium} (New York
  Wiley-Interscience, 1978.~333 p.)

\bibitem[{{Stepnik} {et~al.}(2003){Stepnik}, {Abergel}, {Bernard}, {Boulanger},
  {Cambr{\' e}sy}, {Giard}, {Jones}, {Lagache}, {Lamarre}, {Meny}, {Pajot}, {Le
  Peintre}, {Ristorcelli}, {Serra}, \& {Torre}}]{stepnik03}
{Stepnik}, B., {Abergel}, A., {Bernard}, J.-P., {et~al.} 2003, \aap, 398, 551

\bibitem[{{Stod{\' o}lkiewicz}(1963)}]{stodol63}
{Stod{\' o}lkiewicz}, J.~S. 1963, Acta Astronomica, 13, 30

\bibitem[{{Sugimoto} {et~al.}(2004){Sugimoto}, {Hanawa}, \&
  {Fukuda}}]{sugimoto04}
{Sugimoto}, K., {Hanawa}, T., \& {Fukuda}, N. 2004, \apj, 609, 810

\bibitem[{{Teyssier} {et~al.}(2004){Teyssier}, {Foss{\' e}}, {Gerin}, {Pety},
  {Abergel}, \& {Roueff}}]{teyssier04}
{Teyssier}, D., {Foss{\' e}}, D., {Gerin}, M., {et~al.} 2004, \aap, 417, 135

\bibitem[{{Tilley} \& {Pudritz}(2003)}]{tilley03}
{Tilley}, D.~A. \& {Pudritz}, R.~E. 2003, \apj, 593, 426

\bibitem[{{Warren-Smith} {et~al.}(1985){Warren-Smith}, {Gledhill}, \&
  {Scarrott}}]{warren85}
{Warren-Smith}, R.~F., {Gledhill}, T.~M., \& {Scarrott}, S.~M. 1985, \mnras,
  215, 75P

\bibitem[{{Zaritsky} {et~al.}(1987){Zaritsky}, {Shaya}, {Tytler}, {Scoville},
  \& {Sargent}}]{zaritsky87}
{Zaritsky}, D., {Shaya}, E.~J., {Tytler}, D., {Scoville}, N.~Z., \& {Sargent},
  A.~I. 1987, \aj, 93, 1514

\end{thebibliography}


\end{document}